\documentclass[showpacs,aps,graphicx,twocolumn]{revtex4}%
\usepackage{graphicx}

\begin{document}
\title{Improving the security of multiparty quantum secret sharing against Trojan horse attack}
\author{ Fu-Guo Deng,$^{1,4,5}$\footnote{ E-mail addresses: fgdeng@bnu.edu.cn} Xi-Han Li,$^{1,4,5}$
 Hong-Yu Zhou$^{1,4,5}$  and Zhan-jun Zhang$^{2,3}$\footnote{ Corresponding author.
  E-mail addresses: zjzhang@ahu.edu.cn.} }
\address{$^1$ The Key Laboratory of Beam Technology and Material
Modification of Ministry of Education, Beijing Normal University,
Beijing 100875,
China\\
$^2$ School of Physics and Material Science, Anhui University,
Hefei 230039, China\\
$^3$ Wuhan Institute of Physics and Mathematics, Chinese Academy
of Sciences, Wuhan 430071, China \\
$^4$ Institute of Low Energy
Nuclear Physics, and Department of Material Science and
Engineering, Beijing Normal University, Beijing 100875,
China\\
$^5$ Beijing Radiation Center, Beijing 100875,  China}
\date{\today }

\begin{abstract}
We analyzed the security of the multiparty quantum secret sharing
(MQSS) protocol recently proposed by Zhang, Li and Man [Phys. Rev.
A \textbf{71}, 044301 (2005)] and found that this protocol is
secure for any other eavesdropper except for the agent Bob who
prepares the quantum signals as he can attack the quantum
communication with a Trojan horse. That is, Bob replaces the
single-photon signal with a multi-photon one and the other agent
Charlie cannot find this cheating as she does not measure the
photons before they runs back from the boss Alice, which reveals
that this MQSS protocol is not secure for Bob. Finally, we present
a possible improvement of the MQSS protocol security with two
single-photon measurements and six unitary operations.

\end{abstract}

\maketitle

In classical secret sharing \cite{Blakley}, the boss, say Alice
divides her secret message into two pieces and sends them to her
two agents, Bob and Charlie who are at remote place, respectively,
for her business. Bob and Charlie can reconstruct the secret if
and only if they collaborate. That is $M_A=M_B\oplus M_C$, here
$M_A$, $M_B$ and $M_C$ are the messages hold by Alice, Bob and
Charlie, respectively. The advantage of secret sharing is that one
of the two agents can keep the other one from doing any damage
when they both appear in the process for the business. As
classical signal is in one of the eigenstates of an operator, say
$\sigma_z$, it can be copied freely and fully without leaving a
track. Quantum mechanics provides some novel ways for message
transmitting securely, such as quantum key distribution
\cite{Gisin,longliu,Hwang,BidQKD,ABC}, quantum secure direct
communication \cite{two-step,QSDC,QOTP,QPA}, quantum dense coding
\cite{densecoding,superdense}, and so on.

Quantum secret sharing (QSS) is an important branch of quantum
cryptography \cite{Gisin} and it is the generalization of
classical secret sharing into quantum scenario \cite{HBB99,KKI}.
Since a pioneering QSS scheme was proposed by Hillery, Bu\v{z}ek
and Berthiaume in 1999 by using a three-particle or a
four-particle entangled Greenberger-Horne-Zeilinger (GHZ) state
for sharing a classical information, called HBB99 customarily for
short, there has been a lot of works focused on QSS in both
theoretical
\cite{HBB99,KKI,Bandyopadhyay,Karimipour,guoqss,longqss,delay,
Peng,deng2005,denglongqss,MZ,dengCT,dengmQSTS,zhanglm,cleve} and
experimental \cite{TZG,AMLance} aspects. Almost all the existing
QSS protocols can be attributed to the two types according to
their goals. One is used to distribute a common key among some
users
\cite{HBB99,KKI,Bandyopadhyay,Karimipour,guoqss,longqss,delay,deng2005,denglongqss,MZ,zhanglm},
and the other is used to split a secret including a classical one
\cite{HBB99,KKI,Bandyopadhyay,Karimipour,deng2005} or a quantum
one \cite{Peng,cleve,dengmQSTS,dengCT,zhanglm}.

Recently, Zhang, Li and Man \cite{zhanglm} proposed a multiparty
quantum secret sharing (MQSS) protocol for splitting a classical
secret message among three parties, say Alice, Bob and Charlie
with single photons following some ideas from the Ref.
\cite{QOTP}. In this paper, we will show that the Zhang-Li-Man
MQSS protocol can be eavesdropped by the agent Bob who prepares
the quantum signals with a Trojan horse attack strategy as he can
steal almost all the information encoded by the other agent
Charlie if he replaces the single photon with a multi-photon
quantum signal. As Charlie does not measure the photons for
eavesdropping check before the quantum signal runs back from the
boss Alice, he cannot find this cheating, which is different in
essence to the quantum secure direct communication protocol
\cite{QOTP} and the quantum key distribution protocol with the
practical faint laser pulses \cite{BidQKD}. Moreover, we present a
possible improvement of the Zhang-Li-Man MQSS protocol security
with two single-photon measurements and four unitary operations.

\begin{figure}[!h]
\begin{center}
\includegraphics[width=8cm,angle=0]{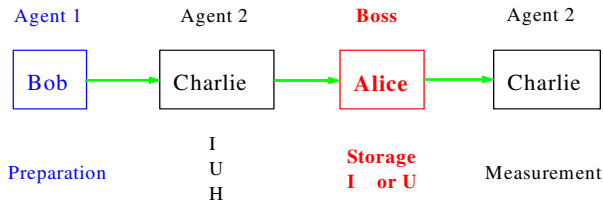} \label{f1}
\caption{ (Color online) Schematic demonstration of the
Zhang-Li-Man multi-party quantum secret sharing protocol
\cite{zhanglm} in the simple case. The two Charlie are the same
agent in two different stages.}
\end{center}
\end{figure}

Let us start with the brief description of the Zhang-Li-Man MQSS
protocol \cite{zhanglm}. We discuss the simple case in which there
are three parties, the Boss Alice, two agents, Bob and Charlie,
shown in Fig.1. The principle for other case is the same as this
simple one with just a little modification or not. In the
Zhang-Li-Man MQSS protocol \cite{zhanglm} the agent Bob prepares a
batch of N single photons with choosing one of the two measuring
bases (MBs), namely, the rectilinear basis, $\sigma_z$, i.e.,
${\vert H\rangle=\vert 0\rangle, \vert V\rangle=\vert 1\rangle}$
and the diagonal basis, $\sigma_x$, i.e., ${\vert
u\rangle=\frac{1}{\sqrt{2}}(\vert 0\rangle + \vert 1\rangle),
\vert d\rangle=\frac{1}{\sqrt{2}}(\vert 0\rangle - \vert
1\rangle)}$ randomly , similar to the Ref. \cite{QOTP}. Bob sends
the single photons to Charlie, and Charlie first takes one of the
three unitary operations ${I, U, H}$ on each photon randomly and
then sends them to Alice. Here $I$ is the identity operation,
$U=i\sigma_y=\vert 0\rangle\langle 1\vert - \vert 1\rangle\langle
0\vert$ whose nice feature is that it flips the state in both
measuring bases \cite{QOTP,BidQKD,zhanglm}, and
$H=\frac{1}{\sqrt{2}}(\vert 0\rangle\langle 0\vert + \vert
1\rangle\langle 0\vert + \vert 0\rangle\langle 1\vert - \vert
1\rangle\langle 1\vert)$ is the Hadamard operation which can
realize the transformation between the two MBs \cite{zhanglm}.
Charlie's operations are equal to the encryptions on the states of
the single photons. After receiving the single photons, Alice
stores most of them and picks out a subset of the photons as the
samples for eavesdropping check. For each sample, Alice requires
Bob publish the initial state first and then Charlie tell her the
encrypting operation, or vice versa. She performs single-photon
measurements on the samples by choosing one of the two measuring
bases $\sigma_z$ and $\sigma_x$ according to the information
published by Bob and Charlie, and analyzes the security of the
transmission of the photons between Bob and Alice. If the error
rate is lower than the threshold, Alice performs the identity
operation $I$ on the state of the single photon if she want to
encode a bit 0, otherwise she performs the operation $U$ on the
photon. Alice sends the single photons to Charlie, and Charlie
will read out the message with the help of Bob's. That is, Bob
tell Charlie the initial states of the photons, and then gets the
operations done by Alice.

As pointed out in the Refs. \cite{HBB99,KKI}, if the dishonest one
in the agents in an MQSS cannot eavesdrop the quantum
communication without disturbing the quantum system, any
eavesdropper can be found out if he wants to steal the
information. In this way, the main goal for the security of an
MQSS is simplified to prevent the dishonest agent from
eavesdropping the information. The Zhang-Li-Man MQSS protocol is
secure if the eavesdropper is Charlie or not the two agents as the
eavesdropping will disturb the quantum system and leave a trick in
the results of the measurements on the sample photons before the
message is encoded on the other photons . Moreover, the parties
can perform a quantum privacy amplification \cite{QPA} on the
batch of the polarized single photons for improving its security
in a noise channel, similar to that in the Ref. \cite{QOTP,QPA}.
However the Zhang-Li-Man MQSS protocol is not secure if the agent
Bob who prepares the quantum signal is the dishonest one as he can
eavesdrop the information freely with Trojan horse attack
\cite{Gisin} even though there are no losses and noise in the
quantum channel. We limit our discussion in this attack below.

For the attack, Bob need only read out the operations that Charlie
encodes on the photons for encrypting their states, and then he
can get the information transmitted by the boss Alice freely. The
Trojan horse attack can be implemented as following: (1) Bob
replaces the single-photon quantum signal with a multi-photon
quantum signal, and Charlie cannot find out this cheating as he
does not measure the photons before they runs back from Alice,
which is different in essence to the quantum secure direct
communication protocol \cite{QOTP}. (2) Bob measures the photons
with some photon number splitters (PNSs: 50/50) and some
detectors. As Bob has the information about the initial states of
the photons, he can read out the operations done by Charlie with a
large probability if there are several photons in each quantum
signal.

\begin{figure}[!h]
\bigskip
\begin{center}
\includegraphics[width=8cm,angle=0]{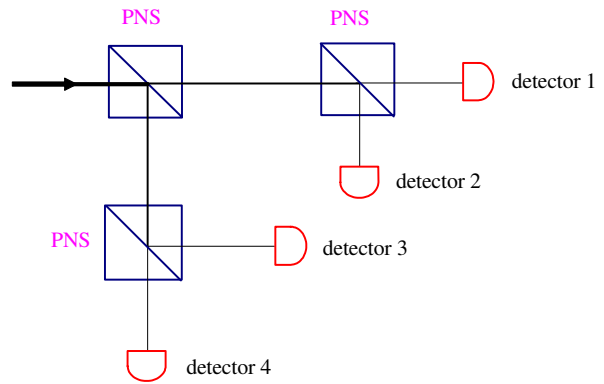} \label{f2}
\caption{ (Color online) The measurements with the photon number
splitters (PNS: 50/50) in the case that there are four photons in
each signal.}
\end{center}
\end{figure}

To present the Trojan horse attack clearly, we use a four-photon
quantum signal as the fake signal with which Bob eavesdrops
Charlie's operations for the sake of simplicity. Bob prepares the
four photons in the same state, say $\vert H\rangle=\vert
0\rangle$ and sends them to Charlie. As Charlie takes his
operation on each signal, the four photons will be performed a
same unitary operation and then there are four copies of the
states encoded by Charlie with the same unitary operation. Bob
uses three PNSs to split the fake signal and performs $\sigma_z$
measurement on each photon, shown in Fig.2. If Charlie takes the
operation $I$ or $U$, the outcomes of the measurements on the four
photons with $\sigma_z$ are the same one with the probability
100\%. On the contrast, Bob gets the same outcome with the
probability $(\frac{1}{2})^4=\frac{1}{16}$ if Charlie operates the
signal with $H$ operation as Bob obtains the outcome "0" or "1"
with the probability 50\% for each photon in the state $\vert
u\rangle=\frac{1}{\sqrt{2}}(\vert 0\rangle + \vert 1\rangle)$. If
Charlie performs the operations $\{I, U, H\}$ on each signal with
the same probability, the probability $P_e$ that Charlie cannot
determine which operation is chosen by Charlie on the signal is
$\frac{1}{3}\times (\frac{1}{2})^4=\frac{1}{48}=2.1\%$. If Bob
replaces the single-photon signal with an N-photon signal, then
$P_e=\frac{1}{3}\times (\frac{1}{2})^N$. When $N=10$,
$P_e=3.26\times 10^{-4}$. Moreover, Bob's eavesdropping does not
introduce errors in the results of the measurements done by
Charlie in the final process if he just measures $N-1$ photons in
the multi-photon signal and sends the other one to Alice.

In the Ref. \cite{QOTP}, as any eavesdropper does not know the
initial state of the quantum signal and the parties take
measurements on a subset of photons chosen randomly for
eavesdropping check of each transmission, the eavesdropper cannot
steal the information with a Trojan horse attack strategy. In the
Zhang-Li-Man MQSS protocol \cite{zhanglm}, there are not those two
features for the agent Charlie. So any eavesdropper can get the
operations done by Charlie with a Trojan horse attack strategy,
including the agent Bob. Certainly, this eavesdropping process is
useful just to Bob, not to other eavesdroppers as they will be
detected by the three legitimate parties if they want to steal the
information about the operations done by Alice or the initial
states prepared by Bob.

For improving the security of the Zhang-Li-Man MQSS protocol
\cite{zhanglm}, the three parties must have the capability to hold
back an eavesdropper to attack the quantum communication with a
Trojan horse. This MQSS protocol is secure if Charlie and Alice
can forbid Bob to eavesdrop the quantum line. Charlie must have
the ability to distinguish whether each quantum signal is a
single-photon one or a multi-photon one before he encrypts the
signals with the unitary operations. For this end, Charlie can
store the quantum signals and chooses randomly a subset of the
quantum signals, similar to Alice. He splits the sample signals
with a photon number splitter (PNS), similar to the Trojan horse
attack done by Bob (see Fig. 2) and then measures the two signals
with the measuring bases $\sigma_z$ and $\sigma_x$ randomly. Just
this modification, Charlie can prevent Bob from eavesdropping
without being detected as both the measurements will have an
outcome if the quantum signal is a multi-photon one. Charlie uses
the probability $P_m$ that there are many photons in each quantum
signal sent by Bob to determine whether Bob is honest. If the
$P_m$ is very low, Bob is a honest one. Certainly, Charlie can
improve the security with three or more PNSs largely, same as that
in Fig.2.

For the symmetry, Charlie can exploit the fourth unitary operation
$\overline{H}=\frac{1}{\sqrt{2}}(\vert 0\rangle\langle 0\vert -
\vert 1\rangle\langle 0\vert - \vert 0\rangle\langle 1\vert -
\vert 1\rangle\langle 1\vert)$ for the encryption. That is,
Charlie chooses randomly one of the four unitary operations $\{I,
U, H, \overline{H}\}$ to encrypt the state of each photon, which
will reduce the probability that Bob obtains the information about
the operations done by Charlie with Trojan horse attack in
particular in the case with a noise quantum channel. That is, the
probability that Bob distinguishes the operations $\{I, U\}$ from
the operations $\{H, \overline{H}\}$ will decrease.

For the integrity, let us describe the modified Zhang-Li-Man MQSS
protocol as follows in brief.

(a) The agent Bob prepares a batch of $N$ single photons $S$
randomly in one of the four polarization states $\{\vert H\rangle,
\vert V\rangle, \vert u\rangle, \vert d\rangle \}$ randomly ,
similar to the Ref. \cite{QOTP}. He sends $S$ to the agent
Charlie.

(b) After receiving $S$, Charlie chooses a sufficiently large
subset of photons in $S$, and splits each signal with a PNS. He
measures the two signals after the PNS with choosing one of the
two MBs $\sigma_z$ and $\sigma_x$ randomly.

(c) Charlie requires Bob to tell him the information about the
original states of the sample photons, and he analyzes the
probability $P_m$ that there are many photons in each quantum
signal sent by Bob. Also he analyzes the error rate $\epsilon_r$
of the sample photons.

(d) If $P_m$ is very low and $\epsilon_r$ is lower than the
threshold, Charlie encrypts almost all the remained photons in $S$
with choosing randomly one of the four unitary operations $\{I, U,
H, \overline{H}\}$. Also, he chooses some samples , say $S_C$ from
the $S$ sequence, and performs them with one of the two operations
$\{\sigma_x, \sigma_z\}$ randomly, and continues to the next step.
Otherwise he discards the results and repeats the quantum
communication from the beginning.

The unitary operations done by Charlie is equivalent to the
uniform encryption on the photons. In a noise channel, the error
correction and the privacy amplification techniques should be used
on those photons for improving the security, same as those in the
Ref. \cite{two-step}. The quantum error correction technique is
not difficult in principle to be implemented, and a quantum
privacy amplification way for the single photons was proposed also
\cite{QPA}. Hence, the quantum communication between Bob and
Charlie can be made secure.

(e) Charlie sends the remained photons, say $S'$ to Alice.  Alice
stores most of the single photons and picks out randomly a
sufficiently large subset of single photons for eavesdropping
check. She tell Bob and Charlie the positions of the sample
photons.  She requires Bob tell her the original states of the
sample photons first, and then Charlie tell her the operations
encoded, or vice versa. For the samples done by Charlie with the
operations $\sigma_x$ and $\sigma_z$, Charlie tell Alice the
positions first and then Alice requires Bob tell her their
original states. Charlie publishes the operations for the samples
$S_C$.

(f) Alice measures each of the sample photons with a correlated
MB, and determines whether there is an eavesdropper monitoring the
quantum line between Bob and Alice, see Fig.1.


(g) If there is no eavesdropper, Alice encodes the photons in $S'$
except for those chosen for eavesdropping check (namely, $S''$),
with the two unitary operations $I$ and $U$ which are coded as the
bits $0$ and $1$, respectively. Surely, Alice should select some
photons from $S''$ as the samples for eavesdropping check and
operate them with $I$ and $U$ randomly, same as
\cite{two-step,QOTP}. Alice sends $S''$ to Charlie.

(h) Charlie reads out the secret message with the help of Bob's.
That is, Bob tell Charlie the original state for each photon, and
then Charlie measures each photon with a correlated MB. Of course,
they should check the eavesdropping of the transmission between
Alice to Charlie before Bob and Charlie cooperate to read out the
message.

The modified Zhang-Li-Man MQSS protocol is in essence equivalent
to the quantum secure direct communication protocol \cite{QOTP}
with quantum privacy amplification \cite{QPA}. So it is secure.
Moreover, it is not difficult to generalize this protocol to the
case with $N$ agents by modifying the original Zhang-Li-Man MQSS
protocol \cite{zhanglm} with the methods discussed above.

In summary, we analyzed the security of the MQSS protocol proposed
by Zhang, Li and Man \cite{zhanglm} and found that this protocol
is secure for any other eavesdropper except for the agent Bob who
prepares the quantum signal as he can attack the quantum
communication with a Trojan horse, i.e., he replaces the original
signal with a multi-photon signal and measures them with some
PNSs. Bob's eavesdropping cannot be detected.  Finally, we present
a possible improvement of the MQSS protocol security with two
single-photon measurements and six unitary operations. With those
modifications, the Zhang-Li-Man MQSS protocol is secure not only
for distributing a common key among the users in MQSS but also for
splitting a secret message, same as the quantum secure direct
communication protocol \cite{QOTP} with the quantum privacy
amplification \cite{QPA}.

\bigskip

Thanks Dr. Su-juan Qin for her helpful discussion. This work is
supported by the National Natural Science Foundation of China
under Grant Nos. 10447106, 10435020, 10254002, A0325401 and
10374010.

\bigskip
\textbf{Addendum}- For improving the security of the Zhang-Man-Li
MQSS protocol completely, the parties of the communication should
have the capability of prevent the agents from eavesdropping. The
eavesdropping done by Charlie who does not know the original
states of the photons prepared by Bob can be detected in the
simple way that Alice requires Charlie tell her the operations
first and then Bob publish the original states of the samples
chosen randomly. For forbidding Bob eavesdrop, the best way may be
that Charlie adds some decoy photons which are prepared by Charlie
and randomly in the states $\vert 0\rangle, \vert 1\rangle, \vert
u\rangle=\frac{1}{\sqrt{2}}(\vert 0\rangle + \vert 1\rangle),
\vert d\rangle=\frac{1}{\sqrt{2}}(\vert 0\rangle + \vert
1\rangle)$, in the sequence $S$. In this way, Charlie should have
an ideal single-photon source.

\end{document}